\documentclass[fleqn,10pt]{wlscirep}
\usepackage[utf8]{inputenc}
\usepackage[T1]{fontenc}
\usepackage{placeins}

\newcommand{\energie}[1]{\textcolor{black}{#1}}

\title{Powering AI at the Edge: A Robust, Memristor-based Binarized Neural Network with Near-Memory Computing and Miniaturized Solar Cell}

\author[1]{Fadi~Jebali}
\author[2]{Atreya~Majumdar}
\author[2]{Cl\'ement~Turck}
\author[2]{Kamel-Eddine~Harabi}
\author[1,4]{Mathieu-Coumba~Faye}
\author[1]{Eloi~Muhr}
\author[1]{Jean-Pierre~Walder}
\author[3]{Oleksandr Bilousov}
\author[3]{Amad\'eo Michaud}
\author[4]{Elisa~Vianello} 
\author[4]{Tifenn~Hirtzlin}
\author[4]{Fran\c{c}ois~Andrieu}
\author[1]{Marc~Bocquet}
\author[2,3]{St\'ephane~Collin}
\author[2,*]{Damien~Querlioz}
\author[1,*]{Jean-Michel~Portal}

\affil[1]{Aix-Marseille Universit\'e, CNRS, Institut Mat\'eriaux Micro\'electronique Nanosciences de Provence, Marseille, France.}
\affil[2]{Universit\'e Paris-Saclay, CNRS, Centre de Nanosciences et de Nanotechnologies, Palaiseau, France.}
\affil[3]{Institut Photovolta\"ique d’Ile-de-France (IPVF),  Palaiseau, France.}
\affil[4]{Universit\'e Grenoble Alpes, CEA, LETI, Grenoble, France.}

\affil[*]{damien.querlioz@c2n.upsaclay.fr, jean-michel.portal@univ-amu.fr}
 
\begin{abstract}
Memristor-based neural networks provide an exceptional energy-efficient platform for artificial intelligence (AI), presenting the possibility of self-powered operation when paired with energy harvesters. However, most memristor-based networks rely on analog in-memory computing, necessitating a stable and precise power supply, which is incompatible with the inherently unstable and unreliable energy harvesters. In this work, we fabricated a robust binarized neural network comprising 32,768 memristors, powered by a miniature wide-bandgap solar cell optimized for edge applications. Our circuit employs a resilient digital near-memory computing approach, featuring complementarily programmed memristors and logic-in-sense-amplifier. This design eliminates the need for compensation or calibration, operating effectively under diverse conditions. Under high illumination, the circuit achieves inference performance comparable to that of a lab bench power supply. In low illumination scenarios, it remains functional with slightly reduced accuracy, seamlessly transitioning to an approximate computing mode. Through image classification neural network simulations, we demonstrate that misclassified images under low illumination are primarily difficult-to-classify cases.  Our approach lays the groundwork for self-powered AI and the creation of intelligent sensors for various applications in health, safety, and environment monitoring.

\end{abstract}
\begin{document}
\maketitle

%
\thispagestyle{empty}


\section*{Introduction}

Artificial intelligence (AI) has found widespread use in various embedded applications such as patient monitoring, building, and industrial safety\cite{cui2018survey}. To ensure security and minimize energy consumption due to communication,  it is preferable to process data at the edge in such systems\cite{warden2019tinyml}. However, deploying AI in extreme-edge environments poses a challenge due to its high power consumption, often requiring AI to be relegated to the  ``cloud'' or the ``fog''\cite{rahmani2018exploiting,qadri2020future}. A promising solution to this problem is the use of memristor-based systems, which can drastically reduce the energy consumption of AI\cite{yu2018neuro,ielmini2018memory}, making it even conceivable to create self-powered edge AI systems that do not require batteries and can instead harvest energy from the environment. Additionally, memristors provide the advantage of being non-volatile memories, retaining stored information even if harvested energy is depleted.

The most-energy efficient memristor-based AI circuits rely on analog-based in-memory computing: they exploit Ohm's and Kirchhoff's laws to perform the fundamental operation of neural networks, multiply-and-accumulate (MAC)\cite{ambrogio2018equivalent,prezioso2015training,wang2018fully}. This concept is challenging to realize in practice due to the high variability of memristors, the imperfections of analog CMOS circuits, and voltage drop effects. To overcome these challenges, integrated memristor-based AI systems employ complex peripheral circuits, which are tuned for a particular supply voltage\cite{xue2021cmos,li2020cmos,yao2020fully,wan202033,jung2022crossbar,khaddam2022hermes,wan2022compute}. This requirement for a stable supply voltage is in direct contrast with the properties of miniature energy harvesters such as tiny solar cells or thermoelectric generators, which provide fluctuating voltage and energy, creating a significant obstacle to realizing self-powered memristor-based AI\cite{ku2015advances}.

In this work, we demonstrate a binarized neural network, fabricated in a hybrid CMOS/memristor process, and designed with an alternative approach that is particularly resilient to unreliable power supply. We demonstrate this robustness by powering our circuit with a miniature wide-bandgap solar cell, optimized for indoor applications. Remarkably, the circuit maintains functionality even under low illumination conditions equivalent to 0.08~suns, experiencing only a modest decline in neural network accuracy. 
When power availability is limited, our circuit seamlessly transitions from precise to approximate computing as it begins to encounter errors while reading difficult-to-read, imperfectly-programmed memristors. 

Our fully digital circuit, devoid of the need for any analog-to-digital conversion, incorporates four arrays of 8,192 memristors each. It employs a logic-in-sense-amplifier two-transistor/two-memristor strategy for optimal robustness, introducing a practical realization of the near-memory computing concept initially proposed in ref.~\cite{bocquet2018,hirtzlin2020digital}. The design is reminiscent of the smaller-scale memristor-based Bayesian machine recently showcased in ref.~\cite{harabi2022memristor}, with the added novelty of logic-in-memory functionality. This feature is achieved by executing multiplication within a robust precharge differential sense amplifier, a circuit initially proposed in ref.~\cite{zhao2014synchronous}. Accumulation is then performed using a straightforward digital circuit situated near-memory. Our system also integrates on-chip a power management unit and a digital control unit, responsible for memristor programming and the execution of fully pipelined inference operations.

We first introduce our integrated circuit and provide a comprehensive analysis of its electrical characteristics and performance across a variety of supply voltages and frequencies. We then characterize the behavior of the circuit under solar cell power, demonstrating its adaptability and resilience even when the power supply is significantly degraded due to low illumination. To further showcase the robustness of the circuit,  we present results from neural network simulations using the popular MNIST and CIFAR-10 datasets. These results highlight the capability of the circuit to perform well even under extremely low illumination conditions.


\section*{Results}

\subsection*{Binarized neural network machine based on distributed memristor modules} 

\begin{figure}[htp]
\centering
\includegraphics[width=0.92\linewidth]{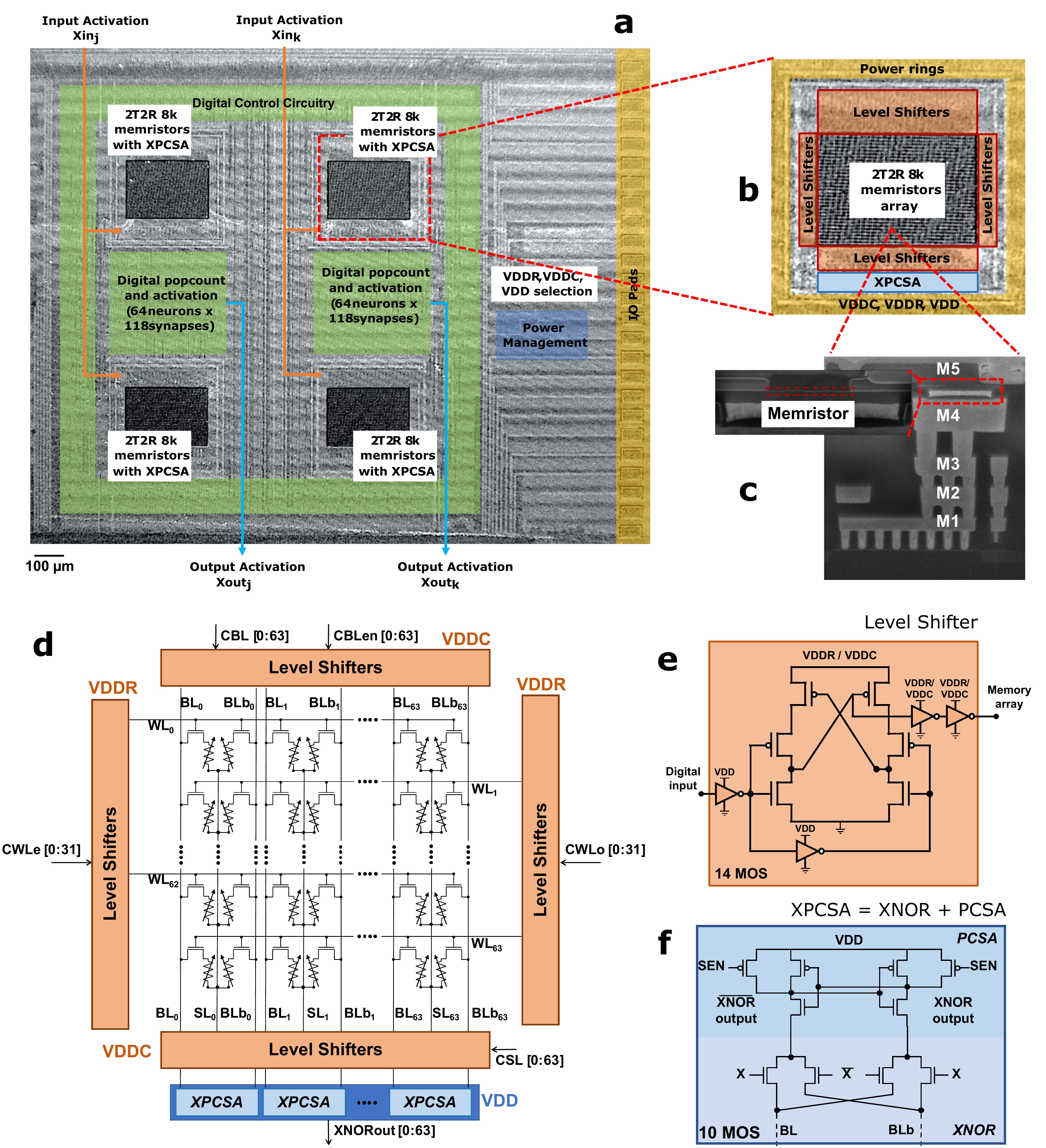}
\caption{
\textbf{Overview of the fabricated memristor-based binarized neural network.}
\textbf{a}~Optical microscopy image of the fabricated die, showing four memory modules and their associated digital circuitry and power management unit.
\textbf{b}~Detail on one of the memory modules.
\textbf{c}~Cross-sectional scanning electron micrograph of a hybrid CMOS/memristor circuit, showing a memristor between metal levels four and five.
\textbf{d}~Schematic of a memory module. For each operation mode, biasing conditions for WL, BL, and SL are given with respect to the power domain (VDDC, VDDR) and VDD.
\textbf{e}~Schematic of the level shifter, used in \textbf{d} for shifting digital voltage input to medium voltages needed during programming operations or nominal voltage during reading operations of the memristors.
\textbf{f}~Schematic of the differential pre-charge sense amplifier PCSA, used to read the binary memristor states, with embedded XNOR function, to compose a XPCSA: it computes an XNOR operation between input activation X and weight (memristor value) during bit-cell sensing.
}
\label{fig:cartoon_system}
\end{figure}

In binarized neural networks, both synaptic weights and neuronal activations assume binary values (meaning $+1$ and $-1$)\cite{courbariaux2016binarized,hubara2016quantized}. These networks are particularly appropriate for the extreme edge, as they can be trained for image and signal processing tasks with high accuracy, while requiring less resources than conventional real-valued neural networks\cite{qin2020binary,zhao2020review}. In these simplified networks, multiplication can be implemented by a one-bit exclusive NOR (XNOR) operation and accumulation by a population count (popcount). The output neuron activations $X_{out,j}$ are, therefore, obtained by
%
\begin{equation}
X_{out,j} = \mathrm{sign} \left( \mathrm{popcount}  \left( XNOR \left( W_{ji},X_{in,i} \right) \right) -T_j \right),
\label{eq:bnn_activation}
\end{equation}
%
using the synaptic weights $W_{ji}$, the input neuron activations $X_{in,i}$  and the output neuron threshold $T_j$.  The quantity \linebreak
$\mathrm{popcount}  \left( XNOR \left( W_{ji},X_{in,i} \right) \right) -T_j$ is a signed integer, referred to as neuron preactivation throughout this paper.

We fabricated a binarized neural network hardware system (Fig.s~\ref{fig:cartoon_system}a,b) employing hafnium-oxide memristors integrated into the back end of a CMOS line to compute equation~\ref{eq:bnn_activation}. The memristors replace vias between metal layers four and five (Fig.~\ref{fig:cartoon_system}c) and are used to  program the synaptic weights and neuron thresholds in a non-volatile manner. 
The system comprises four memristor arrays, each containing 8,192 memristors. These arrays can be used in two distinct configurations: one with two neural network layers featuring 116 inputs and 64 outputs, or an alternative single-layer configuration that has 116 inputs and 128 outputs.
Additionally, we fabricated a smaller die that includes a single 8,192-memristor module with peripheral circuits that provide more flexibility to access memristors. Our circuits use a low-power 130-nanometer process node, which is interesting for extreme-edge applications, as it is cost-effective, offers well-balanced analog and digital performance, and supports a wide range of voltages. Due to the partially academic nature of our process, only five layers of metals are available.

Our design choices aim to ensure the most reliable operation under unreliable power supply and follow the differential strategy proposed in ref.~\cite{hirtzlin2020digital}.  To achieve this, we use two memristors per synaptic weight, programmed in a complementary fashion, with one in a low resistance state and the other in a high resistance state (see Fig.~\ref{fig:cartoon_system}d). We also employ a dedicated logic-in-memory precharge sense amplifier\cite{zhao2014synchronous} to perform the multiplication, which simultaneously reads the state of the two memristors representing the weight and performs an XNOR with its $X$ input (Fig.~\ref{fig:cartoon_system}f). This differential approach makes our circuit highly resilient. It minimizes the effects of memristor variability by ensuring that the sense amplifier functions as long as the memristor in the low resistance state has a lower resistance than the memristor in the high resistance state, even if they deviate significantly from their nominal values. Furthermore, fluctuations in the power supply voltage affect both branches of the sense amplifier symmetrically. This robustness eliminates the need for compensation and calibration circuits, unlike in other analog in-memory computing implementations that require a finely controlled supply voltage.

Our system computes the values of all output neurons in parallel. We provide a detailed description of the pipelined operation of the neural network in Supplementary Note~3, and summarize the main principle here. The neuron thresholds, which are stored in dedicated rows of the memristor arrays, are read simultaneously and transferred to neuron registers located near the memristor arrays. Then,  input neurons are presented sequentially to the memristor array. The accumulation operation of the neural network is performed by integer digital population count circuits that take as input the outputs of the XNOR-augmented sense amplifiers and decrement the neuron registers. These circuits, which are replicated for each output neuron, are located physically near the memristor arrays. This near-memory computing principle saves energy, as only the binarized activations of the output neurons, obtained by taking the sign bit of the threshold register at the end of the inference process, need to be transmitted away from the memories.

As the synaptic weights are stored in non-volatile memory, the system can be turned off and on at any time, cutting power consumption completely, and can immediately perform a new inference or restart a failed one. The programming of the weights needs to be carried out prior to inference, and a forming operation must be performed on each memristor before its first programming operation. A challenge is that the forming operation requires voltages as high as 4.5 volts, whereas the nominal voltage of our CMOS process is only 1.2 volts. To overcome this, we included level shifters  in the periphery circuitry of the memristor arrays (Fig.~\ref{fig:cartoon_system}e), which can sustain high voltages. These circuits, similar to the ones used in ref.~\cite{harabi2022memristor}, use thick-oxide transistors to raise the voltage of the on-chip signals commanding the programming of memristors. The higher-than-nominal voltages are provided by two power pads. Once the memristors have been programmed, these pads can be connected to the digital low-voltage power supply VDD, as high voltages are no longer needed. The details of the memristor forming and programming operation are provided in Supplementary Note~2. Additionally, we incorporated a power management unit and a complete state machine into our fabricated circuit. These components, placed and routed all around the die, are detailed in Supplementary Note~1.

\subsection*{Characterization of the fabricated distributed memory modules BNN machine}

\begin{figure}[htp]
\centering
\includegraphics[width=0.95\linewidth]{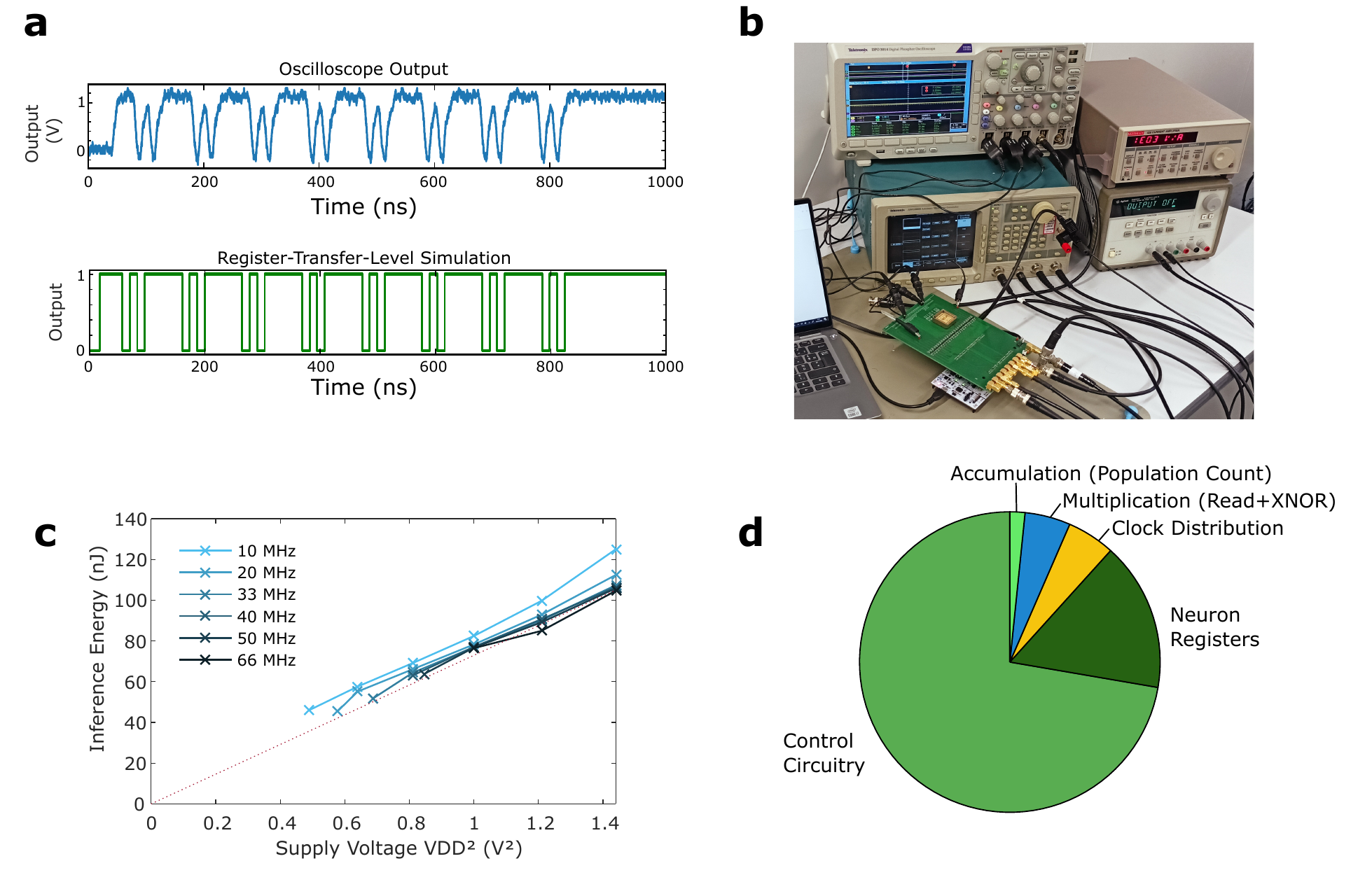}
\caption{
\textbf{Measurements of the memristor-based binarized neural network, employing a lab-bench power supply.}
\textbf{a}~ Sample measurement of the output of the integrated circuit, compared with a delay-less register-transfer level (RTL) simulation.
\textbf{b}~Photograph of the printed circuit board used for the experiments.
\textbf{c}~Measurement of the energy consumption to perform a whole-chip inference, for various operating frequencies and supply voltages.
\textbf{d}~Pie chart comparing the different sources of energy consumption in the system, obtained using simulations (see Methods).
}
\label{fig:system_validation}
\end{figure}

Our fabricated system is functional across a wide range of supply voltages and operating frequencies, without the need for calibration.  As shown in Fig.~\ref{fig:system_validation}a, the measured output of the system, obtained using the setup depicted in Fig.~\ref{fig:system_validation}b, matches the register-transfer-level simulation of our design (see Methods). This first experiment was conducted using the maximum supported supply voltage of our process (1.2~volts) and a clock frequency of 66~MHz.  The energy consumption of the system can be reduced by decreasing the supply voltage, as seen in Fig.~\ref{fig:system_validation}c.
This graph displays the measured energy consumption across various supply voltages and frequencies where the system remained functional. 
The x-axis represents the square of the supply voltage to highlight its direct proportionality to energy consumption: all circuits on-chip, including the sense amplifiers, and with the exception of the power management circuits, function solely with capacitive loads. Notably, energy consumption is largely independent of operation frequency at a given supply voltage. This result, typical for CMOS digital circuits, suggests an absence of short-circuit currents in our design. Supply voltages lower than one volt do not support 66~MHz operation and require slower clock speeds. The lowest measured energy consumption of 45~nJ was achieved at a supply voltage of 0.7 volts (close to the threshold voltage of the transistors in the low-leakage process that we are using) and a clock frequency of 10~MHz. 

Fig.~\ref{fig:system_validation}d details the various sources of energy consumption in our circuit, as determined through simulations based on the process design kit of our technology. (It is not possible to separate the consumption of the different on-chip functions  experimentally.) As the Figure illustrates, a significant portion of the energy is consumed by the on-chip digital control circuitry. In scaled-up systems, this proportion is expected to decrease considerably as the control circuitry would remain largely unchanged. Clock distribution represents only \energie{5.2\%} of the energy, which is lower than typical digital circuits. This is due to the high proportion of circuit area taken up by memristor arrays, which do not require clock distribution. Neuron registers consume a substantial  \energie{16.0\%} of the energy,  owing to their constant activity due to our design decision of not clock-gating them.  This design choice simplified timing constraints in the circuit, ensuring its experimental functionality. However, a fully optimized design would be clock-gated, substantially reducing energy usage for the registers (see Discussion).  The actual multiply-and-accumulate operations, including memristor read with XNOR logic-in-memory and population count, consume a modest  \energie{6.5\%} of the energy.

\begin{figure}[htp]
\centering
\includegraphics[width=0.95\linewidth]{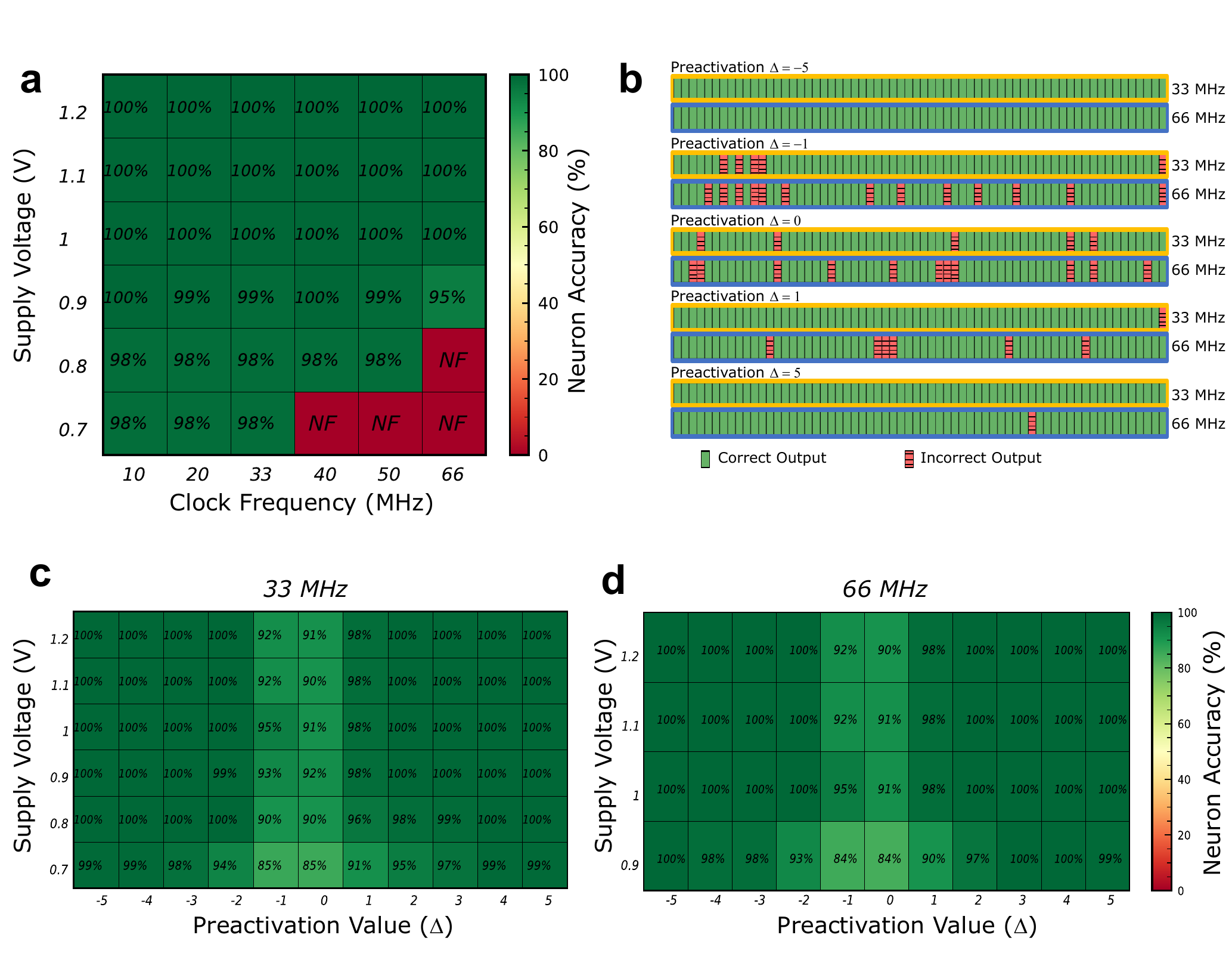}
\caption{
\textbf{Accuracy of the memristor-based binarized neural network.}
\textbf{a}~Measured schmoo plot, presenting mean accuracy of the output neuron activations, for different operation frequency and supply voltage. They were obtained using patterns of weights and inputs chosen to cover all possible neuron preactivations (see Methods). NF means non-functional.
\textbf{b}~Measurements of 64 neurons with preactivations -5, -1, 0, 1, and 5,  at 33 and 66~MHz with a power supply of 0.9~volts. Errors are marked in red.
\textbf{c,d}~Mean accuracy of the output neuron activations, as a function of neuron preactivation $\Delta$ and supply voltage, measured at (\textbf{c}) 33 and (\textbf{d}) 66  MHz (see Methods).
}
\label{fig:error_rates}
\end{figure}

We now present a comprehensive characterization of the accuracy of our fabricated system. Initially, we programmed a memristor array with synaptic weights and neuron thresholds and tested it with neuron inputs, carefully selected to span the entire spectrum of potential output preactivation values (see Methods). Fig.~\ref{fig:error_rates}a presents the measured accuracy (percentage of correct output neurons) across varying supply voltages and operational frequencies in a schmoo plot.  With this setup, we observe no errors when the supply voltage is at least one volt. At 0.9 volts, occasional errors occur at 66 MHz operation, and below this voltage, error rates up to 2\% can manifest at any frequency. We attribute these residual errors to the sense amplifiers, likely due to memristor variability and instability, which cause their resistance to deviate from the target nominal value. Conventional digital circuits incorporating memristors employ strong multiple-error correction codes to compensate for these issues\cite{chang2020envm}.  By contrast, our sense amplifier, owing to its differential nature, can still determine the correct weight even if one memristor exhibits an improper resistance, as long as the memristor programmed in low resistance maintains a lower resistance than the memristor programmed in high resistance. At lower supply voltages, this task becomes more challenging, resulting in the observed residual bit errors.

As neuron errors arise from weight errors, they are only observed when the population count and threshold values of a neuron are comparable. We found that errors were absent experimentally when the difference between the population count and threshold (or neuron preactivation $\Delta$) exceeded five.   
Figs.~\ref{fig:error_rates}c,d, based on extensive experiments (see Methods), depict the error rates for different supply voltages as a function of the neuron preactivation, when the system operates at 33~MHz and 66~MHz. At a supply voltage of 1.2~volts, errors only occur when the preactivation is -1, 0, or 1. At a supply voltage of 0.9~volts, errors are observed for preactivation magnitudes up to five.
To illustrate how errors occur, Fig.~\ref{fig:error_rates} shows measurements of 64 output neurons with varying preactivations values, ranging from -5 to +5, taken at 33 and 66~MHz, with a supply voltage of 0.9~volts.  At this voltage, more errors are observed at 66~MHz than at 33~MHz. Almost all errors detected at 33 MHz continue to exist at 66 MHz. This observation implies that residual errors are likely due to specific weakly-programmed memristors (i.e., complementary memristors programmed with similar resistance), rather than random thermal noise.

\FloatBarrier

\subsection*{Powering the system with harvested energy}

\begin{figure}[htp]
\centering
\includegraphics[width=0.95\linewidth]{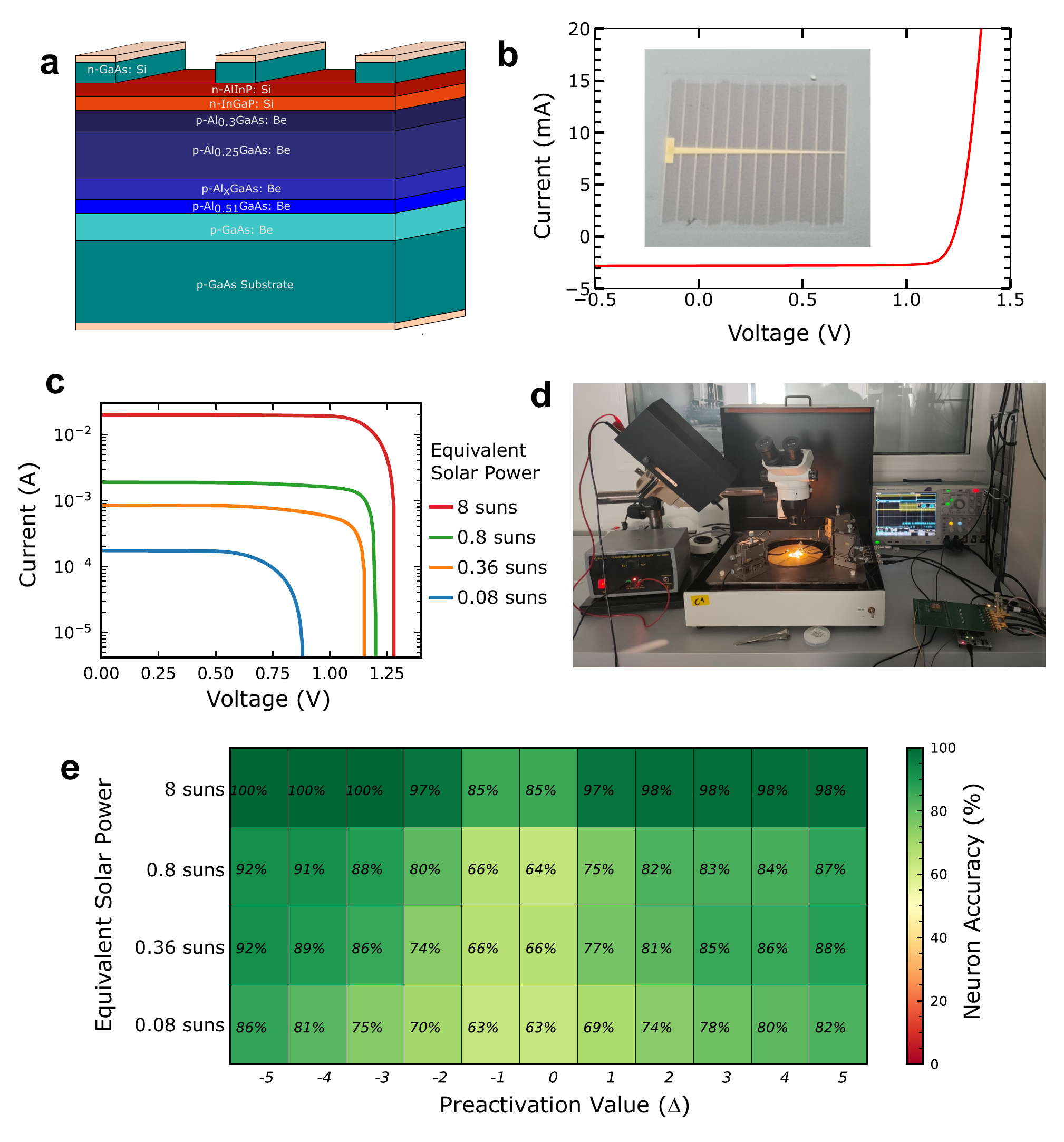}
\caption{
\textbf{Measurements of the binarized neural network powered by a miniature solar cell.}
\textbf{a}~Schematic view of the AlGaAs/GaInP heterostructure solar cell.
\textbf{b}~Photograph of the solar cell, and its measured current-voltage characteristics under one-sun AM1.5 illumination provided by a standardized solar simulator (see Methods).
\textbf{c}~Current-voltage characteristics of the solar cell for various illuminations provided by the halogen lamp (see Methods).
\textbf{d}~Photograph of the experimental setup where the fabricated binarized neural network is powered by the solar cell illuminated by the halogen lamp.
\textbf{e}~Mean measured accuracy of the output neuron activations, with the binarized neural network powered by the solar cell, as a function of neuron preactivation $\Delta$ and solar cell illumination.
}
\label{fig:solar_cell}
\end{figure}

To validate the suitability of our circuit for energy harvesting applications, we connected it to a miniature AlGaAs/GaInP heterostructure solar cell (see Fig.~\ref{fig:solar_cell}a and Methods). Fig.~\ref{fig:solar_cell}b displays a photograph of this cell, along with its current-voltage characteristics measured under standardized one-sun AM1.5 illumination (see Methods).
This type of solar cell, fabricated following the procedure of ref.~\cite{ben20201} (see Methods), with a 1.73~eV bandgap, performs better than conventional silicon-based cells under low-illumination conditions, making it particularly suitable for extreme edge applications. Additionally, due to the wide bandgap, the open-circuit voltage provided by our solar cell (1.23~volts under high illumination) aligns with the nominal supply voltage of our CMOS technology (1.2~volts), unlike silicon solar cells, whose maximum voltage is only 0.7~volts. 

While energy harvesters are typically connected to electronic circuits through intricate voltage conversion and regulation circuits, we demonstrate the resilience of our binarized neural network by directly connecting the power supply pads of our circuit to the solar cell, without any interface circuitry.  In those experiments, the solar cell is illuminated by a halogen lamp (Fig.~\ref{fig:solar_cell}d). 
Fig.~\ref{fig:solar_cell}c presents the current voltage of the solar cell with this setup for various illuminations, expressed as ``equivalent solar powers'' based on the short-circuit current of the solar cell (see Methods). Fig.~\ref{fig:solar_cell}e shows the measured accuracy of our system, plotted as a function of neuron preactivation, similarly to Fig.~\ref{fig:error_rates}d. 

Under an equivalent solar power of 8~suns, the circuit performs almost equivalently to when powered by a 1.2~volts lab bench supply. When illumination decreases, even under a very low equivalent solar power of 0.08~suns where the characteristics of the solar cell is strongly degraded, the circuit remains functional. However, its error rate increases, especially for low-magnitude preactivation values. The circuit naturally transitions to an approximative computing regime: neurons will large-magnitude preactivations are correctly computed, but those with low-magnitude preactivations may exhibit errors.

\begin{figure}[htp]
\centering
\includegraphics[width=0.82\linewidth]{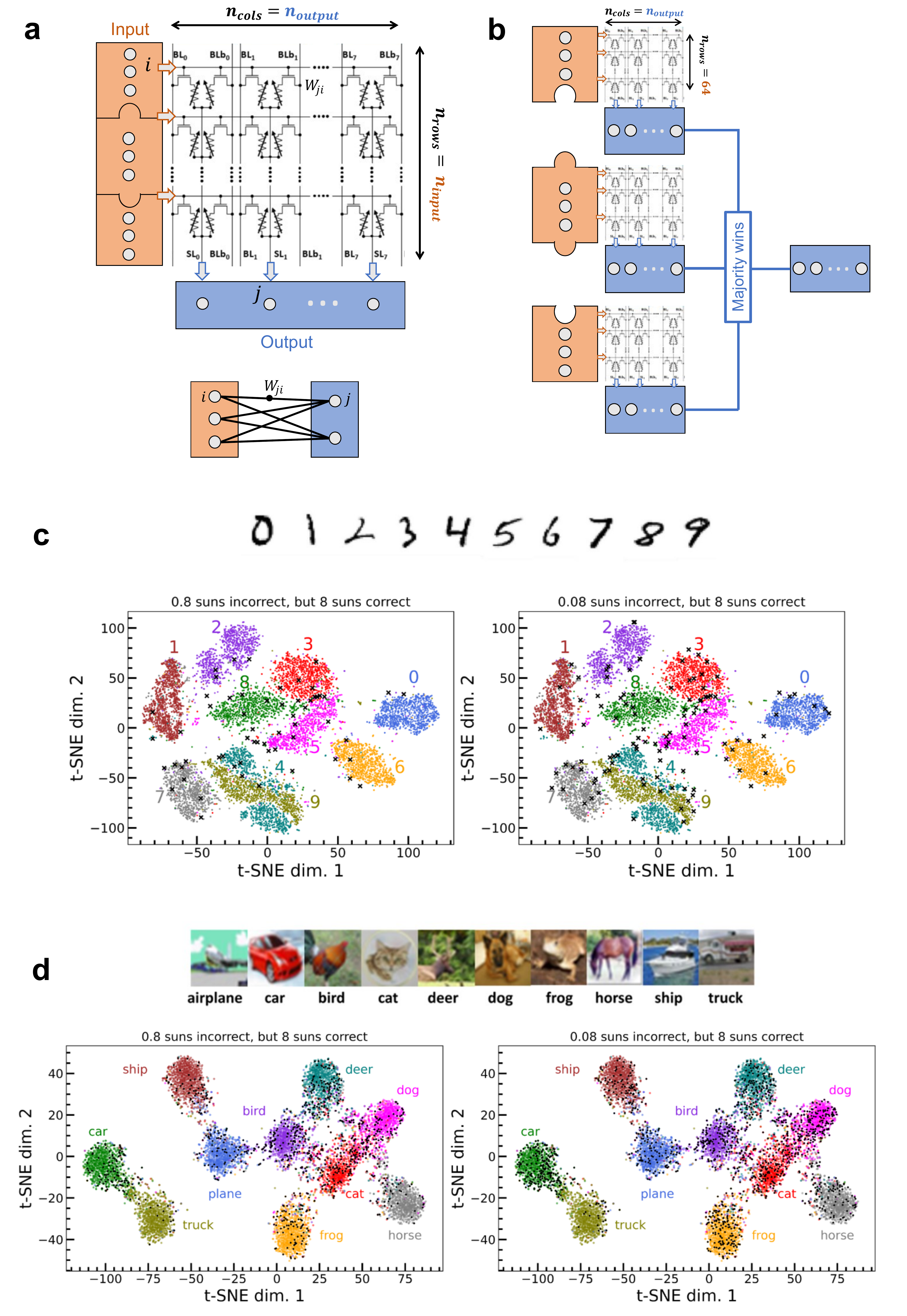}
\caption{
\textbf{Neural-network-level investigations.}
\textbf{a,b}~Illustration of our method for mapping arbitrarily-shaped binarized neural networks to 64$\times$128 memristor arrays. The detailed methodology is presented in Supplementary Note~4.
\textbf{c}~t-distributed stochastic neighbor embedding (t-SNE) representation of the MNIST test dataset. The datapoints incorrectly classified under 0.8~suns (left) and 0.08~suns (right) equivalent illumination, but which would be correctly classified under 8~suns, are marked in black. These results are obtained using a binarized fully-connected neural network (see Methods).
\textbf{d}~Same graphs as  \textbf{c}, obtained for the CIFAR-10 dataset and using a convolutional neural network (see Methods).
}
\label{fig:neuralnets_simul}
\end{figure}

\begin{table}[htp]
\begin{center}
\begin{tabular}{cccc}
\hline
      Equivalent Solar Power     & MNIST Accuracy& CIFAR-10 Accuracy \\
\hline
Baseline    &  97.2\% &  86.6\%  \\
8 suns    &  97.1\% &  83.6\%  \\
0.8 suns    &  96.9\% &   78.2\%  \\
0.36 suns    &  96.9\% &   78.3\%  \\
0.08 suns    &  96.5\% &   73.4\%  \\
 \hline
\end{tabular}
\caption{
Simulated accuracy of solar-cell power a fully-connected (MNIST task) and a convolutional (CIFAR-10 task) binarized neural network under various illuminations. The software baseline assumes no bit error (see Methods).
}
\label{table:perf_neurnets}
\end{center}
\end{table}

We now evaluate the performance of our circuit on neural networks. Our system functions with 128$\times$64 memristor arrays; however, in practice, neural networks can have various structures. To map neural networks to our hardware, we employ a technique that subdivides neural network layers into several binarized arrays and then obtains the value of output neurons through majority votes of the binary output of each array  (see Figs.~\ref{fig:neuralnets_simul}a,b). This method, which we describe in more detail in Supplementary Note~4, is highly efficient in terms of hardware usage and causes only moderate accuracy degradation compared to software-based neural networks on the two tasks considered here: Modified National Institute of Standards and Technology (MNIST) handwritten digit recognition and CIFAR-10 image recognition. 

To evaluate the classification accuracy of our hardware, we incorporated the error rates measured experimentally as a function of preactivation value and illumination (Fig.~\ref{fig:solar_cell}d) into neural network simulations (see Methods). Table~\ref{table:perf_neurnets} lists the obtained accuracy on a fully-connected neural network trained on MNIST and a convolutional neural network trained on CIFAR-10 (see Methods). Remarkably, the MNIST accuracy is hardly affected by the bit errors in the circuit: even under very low illumination equivalent to 0.08~suns, the MNIST accuracy drops by only 0.7 percentage points. Conversely, bit errors significantly reduce the accuracy of the more demanding CIFAR-10 task. Under 0.08~suns, the accuracy drops from the software baseline of 86.6\% to 73.4\%. The difference with the MNIST  arises because more neurons tend to have low-magnitude preactivation when solving CIFAR-10, as the differences between classes are more subtle.

To further understand the impact of low illumination on neural network performance, we plotted the t-distributed stochastic neighbor embedding\cite{van2008visualizing} (t-SNE) representation of the MNIST test dataset in Fig.~\ref{fig:neuralnets_simul}b. This technique represents each image as a point in a two-dimensional space, where similar images cluster together and dissimilar ones reside at a distance. In the left image, we marked in black the images that were correctly classified by a neural network under illumination equivalent to 8~suns, but incorrectly under 0.8~suns. Interestingly, these images tend to be on the edges of the clusters corresponding to the different digit classes, or even outliers that do not belong in a cluster. This suggests that the images that the network starts misclassifying under 0.8~suns tend to be subtle or atypical cases. The right image shows that this effect intensifies under illumination equivalent to 0.08~suns, with a few images inside clusters also being misclassified. Fig.~\ref{fig:neuralnets_simul}c presents the same analysis for the CIFAR-10 dataset. The trend of incorrectly classified images under low illumination tending to be edge or atypical cases persists, albeit less pronounced than with MNIST.

\FloatBarrier

\section*{Discussion}

Our circuit exhibits an original behavior when solving tasks of varying difficulty levels. For simpler tasks such as MNIST, the circuit maintains accuracy even when energy is scarce. When addressing more complex tasks, the circuit becomes less accurate as energy availability decreases, but without failing completely.  This self-adaptive approximate computing feature has several roots and can be understood by the circuit's memory read operations. They are highly robust due to their differential nature: fluctuations of the power supply affect both branches of the sense amplifier equally. Still, when power voltage fluctuates or becomes low, some memory reads fail. Nevertheless, binarized neural networks are highly robust to weight errors, which in many cases do not change neuron activation\cite{hirtzlin2019outstanding,buschjager2021margin}. Even in the worst case, weight errors cause some images to be misclassified, but these are typically atypical or edge cases. Therefore, when the power supply degrades, the AI naturally becomes less capable of recognizing harder-to-classify images.

In this context of low-quality power supply, memristors offer distinct advantages over conventional static RAMs. While static RAMs lose stored information upon power loss, memristors retain data.  Furthermore, when the supply voltage becomes low, static RAMs are prone to read disturb, meaning that a read operation can change the bit stored in a memory cell. In contrast, memristors exhibit near-immunity to read disturb effects, especially when read by precharge sense amplifiers\cite{harabi2022memristor} (we observed no read disturb in our experiments), and are non-volatile (ten-years retention has been demonstrated in hafnium-oxide memristors\cite{golonzka2019non}).

After eliminating the energy used by the digital control circuitry (finite state machine), our circuit has an energy efficiency of \energie{2.9} tera-operations per second and per watt (TOPS/W) under optimal conditions (10~MHz frequency, supply voltage of 0.7~volts). By further subtracting the energy consumption of clock distribution and neuron registers that can be eliminated through clock gating, and simultaneously optimizing the read operation (see Methods), energy efficiency increases to \energie{22.5}~TOPS/W. Due to the digital nature of our circuit, this number would scale favorably if a more current CMOS process was used.  For example, employing the physical design kit of a fully-depleted silicon-on-insulator 28-nanometer CMOS process, we found that the energy efficiency of a clock-gated design would reach 397 TOPS/W (see Methods). Supplementary Note~5 compares these numbers and other properties of our digital system with fabricated emerging memory-based analog in-memory computing circuits. The most noteworthy comparison is with a recent study that presents an analog magnetoresistive memory (MRAM) based 64x64 binarized neural network fabricated in a 28-nanometer process\cite{jung2022crossbar}, which has a measured energy efficiency of 405 TOPS/W, which surpasses our projection  slighly. However, this energy efficiency comes with the need for complex compensation and calibration circuits, matched to a stable power supply, which is not suitable with the unreliable power supply delivered by energy harvesters.

Our circuit can function with power supplies as low as 0.7~volts, enabling us to power it with a wide-bandgap solar cell optimized for indoor applications, with an area of only a few square millimeters, even under low illumination equivalent to 0.08~suns. Such lightweight, ultrathin solar cells can also be transferred into a fully-integrated, self-powered device\cite{chen201919,massiot2020progress}.  Supply voltages lower than 0.7~volts result in significant inaccuracies in memristor readings due to the high threshold voltages of the thick-oxide transistors in our process. Employing a process with a lower threshold voltage thick-oxide transistor option could enable operation at lower supply voltages, broadening compatibility with various solar and non-solar energy harvesters. Some very low-voltage harvesters (e.g., thermoelectrics) may still require the voltage to be raised, which can be accomplished on-chip using switched capacitor circuits like Dickson charge pumps\cite{yoon2018area}. Self-powered AI at the edge, therefore, offers multiple opportunities to enable the development of intelligent sensors for health, safety, and environmental monitoring.

\section*{Acknowledgements}
This work received funding within the ECSEL Joint Undertaking (JU) project storAIge in collaboration with the European Union’s H2020 research and innovation program and National Authorities, under grant agreement numbers 101007321. This work was also supported by European Research Council starting grant NANOINFER (reference: 715872), by the Agence Nationale de la Recherche through the NEURONIC (ANR-18-CE24-0009) grant, by the French Government in the framework of the ``Programme d’Investissement d’Avenir'' (ANR-IEED-002-01), and with the support of the cleanroom RENATECH network. It also benefits from a France 2030 government grant managed by the French National Research Agency (ANR-22-PEEL-0010). The authors would like to thank J.~Grollier and L.~Hutin  for discussion and invaluable feedback. Parts of this manuscript were revised with the assistance of a large language model (OpenAI ChatGPT).

\section*{Author contributions statement}
J.M.P designed the hardware binarized neural network, using a flow developed with J.P.W., and with contributions from M.C.F., E.M., and D.Q. The system was fabricated under the direction of E.V. and F.A. F.J. performed the on-chip experimental measurements, under the direction of M.B. C.T., and K.E.H. analyzed the energy consumption of the system. A.M. performed the neural network-level analyses. O.B., A.M., and S.C. developed and fabricated the solar cells. T.H. developed the binarized neural network simulator. D.Q. and J.M.P. directed the project and wrote the initial version of the manuscript. All authors discussed the results and reviewed the manuscript. 

\section*{Competing interests}
The authors declare no competing interests.

\section*{Data availability}
The datasets analyzed and all data measured in this study are available from the corresponding author upon reasonable request.

\section*{Code availability}
The software programs used for modeling the Binarized Neural Network machine are available from the corresponding author upon reasonable request.

\section*{Methods}
 
\subsection*{Fabrication of the demonstrator}

The MOS part of our demonstrator was fabricated using a low-power 130-nanometer foundry process up to the fourth layer of metal. Memristors, composed of a TiN/HfO$_x$/Ti/TiN stack, were then fabricated on top of exposed vias. The active 10-nanometer thick HfO$_x$ layer was deposited by atomic layer deposition. The Ti layer is also 10-nanometer thick, and the memristor structure has a diameter of 300 nanometers. A fifth layer of metal was deposited on top of the memristors. 25 input/output pads are aligned to be compatible with a custom probe card. A packaged version of the demonstrator was also assembled in a J-Leaded Ceramic Chip Carrier with 52 leads. 

\subsection*{Design of the demonstrator}

The memristor-based Binarized Neural Network is a hybrid CMOS/nanotechnology integrated circuit with distributed memory modules within the logic. The design of the memory module includes the array and peripheral circuits, such as the XNOR-augmented precharge sense amplifiers (Fig.~\ref{fig:cartoon_system}b) and the level shifter circuits (Fig.~\ref{fig:cartoon_system}c). The memory modules were designed using a full-custom flow under the Cadence Virtuoso electronic design automation (EDA) tool and were simulated using the Siemens Eldo simulator. Verification steps, i.e., layout versus schematic check and design rule check, were performed using Calibre tools.

The level shifter circuit (Fig.~\ref{fig:cartoon_system}e) was designed with thick-oxide MOS transistors supporting up to five volts.
To isolate the precharge current sense amplifier during the forming or programming operations, the four XNOR MOS transistors (the ones connected to the input X in Fig.~\ref{fig:cartoon_system}f) were designed with thick gate oxide. The sense amplifier itself was constructed using thin gate oxide transistors. The memory modules architecture also includes four dedicated power rings: one for VDDR, one for VDDC, one for VDD, and one for the ground (GND). An abstract view of the memory modules was generated using the Cadence abstract generator. The power switch unit, which has to sustain up to 4.5 volts during the forming operation, was also designed using thick-oxide transistors, following the same full-custom flow as the memory modules.  A Liberty Timing Files (.lib)  related to the abstract view of the full custom blocks was handwritten and  a Synopsys database file (.sdb) was generated using the Synopys Library Compiler. 

The overall machine core follows a digital on-top flow, where all digital blocks (e.g., controller logic, population count decounter, neuron registers) are described using the VHSIC Hardware Description Language (VHDL), including the full custom blocks entity, synthesized using the Synopsys Design Compiler, and finally placed and routed, including the full-custom abstract view, using the Cadence Encounter RTL-to-GDSII tool, following a semi-automated flow developed by the foundry. All digital circuits use thin-oxide high-threshold transistors and are biased to VDD. Logical verification of the core, including the memory modules, described with an equivalent handmade VHDL behavioral description, and the power switch, described with an equivalent handmade VerilogA description, were performed using Siemens Questa mixed-signal simulator. The memory modules equivalent VHDL description and the power switch VerilogA equivalent descriptions were first assessed against their electrical schematic counterparts, simulated with Siemens Eldo electrical simulator. The connection of the machine layout to the 25 input/output pads was accomplished manually in a full-custom fashion.

Supplementary Note~1 describes the digital control circuitry and the power management unit with more technical details. Supplementary Note~2 details the methodology used by our circuit for forming and programming the memristors. Supplementary Note~3 lists the steps of the pipelined inference operation of the circuit.

\subsection*{Fabrication of the solar cell}

The fabrication of solar cells in this study was carried out according to the procedures described in ref.~\cite{ben20201}. The semiconductor stack was grown on a GaAs substrate using molecular beam epitaxy and consisted of the following sequence of layers: p-GaAs:Be (300 nm), p-Al$_{0.51}$GaAs:Be (50 nm), p-AlGaAs:Be with a linear gradient from 51\% to 25\% Al (100 nm), p-Al$_{0.25}$GaAs:Be (1900 nm), n-Al$_{0.3}$GaAs:Si (100 nm), n-InGaP:Si (50 nm), n-AlInP:Si (20 nm), and n-GaAs:Si (300 nm) (see Fig.~\ref{fig:solar_cell}a).

The front metal grid was defined using standard photolithography techniques, followed by metal evaporation (NiGeAu) and lift-off processes. Wet chemical etching was used to separate the mesa structures of the different cells, and to etch the top 300 nm-thick GaAs contact layer outside the front grid area. The back contact (TiAu) is deposited on the backside of the substrate. No anti-reflection coating was added. The size of the solar cells is $5$ mm $ \times 5\rm$ mm.

\subsection*{Measurements of the system with lab-bench power supply}
  
The measurements of our system were conducted on the packaged version. The binarized neural network integrated circuit is mounted on a dedicated printed circuit board (PCB) featuring level shifters and SubMiniature A (SMA) connectors (see Fig.~\ref{fig:system_validation}b). The PCB connects the different input and output signals of the packaged chip to an STM32F746ZGT6 microcontroller unit,  a Tektronix AWG2005 arbitrary waveform generator, and a Tektronix DPO 3014 oscilloscope. The voltage for the level shifters of the PCB is supplied by an Agilent E3631A power supply. The microcontroller unit is connected to a computer using a serial connection, while lab-bench equipments are connected to the computer using a National Instruments GPIB connection. The whole setup is controlled using python within a single Jupyter notebook.

Supplementary Note~2 details the memristor forming and programming operations, and we summarize them here. Before starting any measurement, all the memristors are formed, sequentially, under the control of the on-chip digital control block (Supplementary Note~{1}). During this operation, the VDDC supply voltage is set to 4.5~volts, VDDR to 2.7~volts, and VDD to 1.2~volts, during ten microseconds. After this initial forming step, the memristor array is programmed with the desired pattern (synaptic weights and neuron thresholds). The programmed data are transmitted to the microcontroller unit, which sends them to the binarized neural network integrated circuit row-by-row. To program a memory cell to HRS, the digital control block connects VDDC to 2.7 volts and VDDR to 4.5 volts, with VDD fixed at 1.2 volts. To program a memory cell to LRS, VDDC, and VDDR are both connected to 2.7 volts, and VDD is connected to 1.2 volts. The two memristors of each bit cell are always programmed in a complementary fashion (i.e., either LRS/HRS or HRS/LRS). The digital circuitry controls the programming operations based on the weight value for each memristor, and applies the programming pulses during six microseconds.

To perform inference (see Supplementary Note~3), input neuron activations are sent through the microcontroller unit for each row, and the output of the integrated circuit is captured by the microcontroller unit and stored in a comma-separated values (CSV) file. 
To obtain the schmoo plots shown in Fig.~\ref{fig:error_rates}, the power supply voltage VDD was varied from 1.2 to 0.7 volts, for all considered operation frequencies. The saved outputs of the integrated circuit were compared to the expected outputs to extract the system's accuracy.   

To measure the power consumption of the circuit (Fig.~\ref{fig:system_validation}c), the VDD power supply of the test chip is connected to a Keithley 428 current amplifier. The output of the Keithley 428 is connected to the oscilloscope to obtain the current during inference.

\subsection*{Measurements of the system powered by the solar cell}

We first characterized the current-voltage characteristics of the solar cell (Fig.~\ref{fig:solar_cell}b),  using a certified solar simulator providing a one-sun  (100~mW/cm$^2$) AM1.5 illumination.  To power our binarized neural network by the solar cell,  we switched to a more accessible variable-illumination halogen lamp  (Fig.~\ref{fig:solar_cell}c), whose spectrum does not match   AM1.5 solar light. To obtain an equivalent solar power, we measured the current-voltage characteristics of the solar cell under this lamp (Fig.~\ref{fig:solar_cell}b) using the source measure unit mode  of a Keysight B1530A unit. We calculate the equivalent solar power by dividing the short circuit current by the one under  one-sun AM1.5 illumination.

We then directly connected our binarized neural network to the solar cell and conducted inference measurements using the same methodology as with the lab-bench power supply. To accomplish this, we connected all three power pads of the circuit (VDD, VH, VM, see Supplementary Note~1) to the solar cell, as high supply voltages are not needed to perform inference. To obtain the  schmoo of Fig.~\ref{fig:solar_cell}e, we used the same methodology as for Fig.~\ref{fig:error_rates}, but by varying the halogen light illumination instead of the bench power supply voltage.

\subsection*{Energy consumption estimates}

Energy measurements of the system (shown in Fig.~\ref{fig:system_validation}c) cannot differentiate the consumption of the different elements of the circuit, as they all share the same power supply. To overcome this limitation (as illustrated in Fig.~\ref{fig:system_validation}s), we relied on computer simulations of our circuit using commercial integrated circuit design tools. 

We obtained energy estimates during the inference phase, after the memristors were formed and the memory programmed. The consumption of the memristor arrays was determined using circuit simulations (based on the Siemens Eldo simulator), which also accounted for parasitic capacitance extracted from the memristor array layout. For the remainder of the system, we analyzed it using the Cadence Voltus power integrity solution framework on the placed-and-routed design, incorporating all parasitics. We utilized a value change dump (VCD) file obtained from a test bench simulation to ensure a realistic situation.

The memristor array blocks are full-custom and, therefore, not included in the standard library of the foundry. This raised a concern regarding the continuous flow of values before and after the memristor array when performing the energy analysis. To address this, we wrote a new liberty (.lib) file, specifically for use in the energy analysis, based on the actual output values during simulation to ensure that the flow before and after the memory was respected during the inference phase. 

In our fabricated circuit, the neuron registers are enabled when an XNOR-augmented sense operation is performed. We chose not to clock-gate these registers to avoid any timing risk in our test chip; however, this strategy can be employed to reduce the energy consumption of a final design. Therefore, we also designed a clock-gated version of our circuit and estimated its energy consumption using the same flow as for the fabricated version. This clock-gated version also uses an optimized read process requiring fewer clock cycles. We finally estimated the energy consumption of a scaled-down version of the design in a commercial 28-nanometer fully-depleted silicon-on-insulator CMOS design kit. For this analysis, the memristor array was entirely redesigned in the 28-nm design kit. For the digital part, we use a  scaling factor  relating the typical energy consumption of equivalent circuits in the two commercial technology nodes.

\subsection*{Neural-network level investigations}

For the neural network simulations presented in Fig.~\ref{fig:system_validation}, we used a fully connected  architecture for the MNIST handwritten digit recognition task, and a convolutional neural network architecture for the CIFAR-10 image classification task. Except for the input to the first layer, the activations and weights of the network were binarized, following the binarized neural network implementation \cite{hubara2016quantized}. The fully connected (FC) network had two hidden layers with 1,102 and 64 neurons, whereas the convolutional architecture was based on the VGG-16 network, and it consisted of 3x3 kernels for convolutions (Conv), batch normalizations (BN), and nxn for MaxPool (MPn) and reads: [Conv~198, BN, Conv~198, MP~2, BN, Conv~354, BN, Conv~354, MP2, BN, Conv~738, BN, Conv~406, MP3, FC(1102-1102-10)]. The number of hidden layer units and convolutional filters were chosen in accordance with the dedicated mapping technique described in Supplementary Note~4, such that the total number of blocks is always odd when a block size of 58 is used. 

We trained the networks without errors and with the mapping technique implemented. The input neurons of the first layer and the output neurons of the final layer are non-binary, so we did not include circuit-induced errors in these layers, as they require different circuits. The convolutional network was trained for 500 epochs with the Adam optimizer with weight decay and  a cosine annealing learning rate scheduler. The fully-connected network was trained with the same optimizer for 200 epochs with a step learning rate scheduler \cite{loshchilov2016sgdr}. Only after the training was completed were the errors introduced during the inference step, using a dedicated Pytorch code reproducing the error rate measured experimentally (Fig.~\ref{fig:solar_cell}e). The error rate of the circuit for a certain level of illumination and a certain preactivation $\Delta$ was taken as the probability of having an error in the neuronal output. The PyTorch deep learning framework was used to perform all the neural network simulations. 
\bibliography{sample}
\end{document}